 \documentclass[preprint2]{aastex}

\usepackage{epsfig}
\usepackage{epstopdf}
\usepackage{graphicx}

\newcommand{\Mj}{M$_\mathrm{Jup}$}

\slugcomment{Submitted to ApJL, Accepted}

\shorttitle{Confirmation of the planet around HD\,95086}
\shortauthors{Rameau et al.}

\begin{document}


\title{Confirmation of the planet around HD\,95086 by direct imaging}


\author{J. Rameau\altaffilmark{1}, G. Chauvin\altaffilmark{1}, A.-M. Lagrange\altaffilmark{1}, T. Meshkat\altaffilmark{2}, A. Boccaletti\altaffilmark{3},  S.P. Quanz\altaffilmark{4}, T. Currie\altaffilmark{5}, D. Mawet\altaffilmark{6}, J.H. Girard\altaffilmark{6}, M. Bonnefoy\altaffilmark{7}, \& M. Kenworthy\altaffilmark{2}}

\affil{
\altaffilmark{1}UJF-Grenoble 1 / CNRS-INSU, Institut de Plan\'etologie et d'Astrophysique de Grenoble (IPAG) UMR 5274, Grenoble, F-38041, France\\
\altaffilmark{2}Leiden Observatory, Leiden University, P.O. Box 9513, 2300 RA Leiden, the Netherlands\\
\altaffilmark{3}LESIA, Observatoire de Paris, CNRS, University Pierre et Marie Curie Paris 6 and University Denis Diderot Paris 7, 5 place Jules Janssen, 92195 Meudon, France\\
\altaffilmark{4}Institute for Astronomy, ETH Zurich, Wolfgang-Pauli-Strasse 27, 8093 Zurich, Switzerland\\
\altaffilmark{5}Department of Astronomy and Astrophysics, University of Toronto, 50 St. George St., Toronto, Ontario, M5S 1A1, Canada\\
\altaffilmark{6}European Southern Observatory, Casilla 19001, Santiago 19, Chile\\
\altaffilmark{7}Max Planck Instiute f\"ur Astronomy, K\"onigsthul 17, D-69117 Heidelberg, Germany\\
}

\altaffiltext{}{Electronic adress: julien.rameau@obs.ujf-grenoble.fr\\
Based on observations collected at the European Organisation for Astronomical Research in the Southern Hemisphere, Chile, under programs number 291.C-5023.}


\begin{abstract}
VLT/NaCo angular differential imaging at L\,' ($3.8~\mu m$) revealed a probable giant planet comoving with the young and early-type HD\,95086 also known to harbor an extended debris disk. The discovery was based on the proper motion analysis of two datasets spanning 15 months. However, the second dataset suffered from bad atmospheric conditions, which limited the significance of the re-detection at the $3\sigma$ level. 
In this Letter, we report new VLT/NaCo observations of HD 95086
obtained on 2013 June 26-27 at L\,' to recover the planet candidate.  We
unambiguously redetect the companion HD\,95086\,b with multiple independent
pipelines at a signal-to-noise ratio greater than or equal to 5.  Combined
with previously reported measurements, our astrometry decisively shows
that the planet is comoving with HD 95086 and inconsistent with a
background object. With a revised mass of $5\pm2$ Jupiter masses, estimated from its L\,' photometry and ``hot-start" models at $17\pm4~$Myr, HD\,95086\,b becomes a new benchmark for further physical and orbital characterization of young giant planets.
\end{abstract}





\section{Introduction \label{sec:intro}}
We reported in \citet{rameau13b} the discovery of a probable $4-5$~\Mj~giant planet at $\sim56~$AU (projected separation) from its host-star HD\,95086 (A8, $90.4$~pc). The star belongs to the $17\pm4~$Myr old Lower Centaurus Crux association \citep{pecaut12, meshkat13b}.
Based on two observing epochs acquired in January 2012
 and March 2013, we showed that the detected point-source was likely
 comoving with HD\,95086. However, the confirmation remained ambiguous
 owing to poor redetection at the $3\sigma$ level in March
 2013. Additional observations at Ks-band ($2.18~\mu m$), and even more
 recently in H-band ($1.6~\mu m$) with Gemini/NICI \citep{toomey03} indicate very red colors, Ks-L\,'$\ge1.2~$mag
 and H-L\,'$\ge 3~$mag \citep{rameau13b, meshkat13b}. These colors are compatible with a cool and dusty planetary
 atmosphere. This allowed us to further reject a contamination by a background
 source. The companion L' apparent flux and colors upper limit are
 compatible with the predictions of the ``hot-start" DUSTY evolutionary models \citep{chabrier00} whose color and absolute predictions were recomputed using BT-Settl atmospheric models \citep{allard12} for a mass
 below $5~$\Mj.
 
We re-observed the system on June-26th and 27th using VLT/NaCo with the aim of 
 redetecting HD\,95086\,b with a high level of confidence and to constrain its proper motion relative to HD\,95086. 
  




\section{Observing strategy and image processing\label{sec:obs}}
\subsection{Observations}

\begin{table*}[t!]
\caption{Observing log of HD\,95086 with VLT/NaCo. \label{tab:log}}
\centering
\scriptsize
\begin{tabular}{llllllllll}     
\tableline
\tableline
Date & Cam./Filter & DIT $\times$ NDIT & N$_{\rm exp}$ & $\pi$-start/end& $\langle$Airmass$\rangle$\tablenotemark{a} & $\langle\varpi\rangle$\tablenotemark{a} & $\langle\tau_0\rangle$\tablenotemark{a} & $\langle\mathrm{E}_\mathrm{c}\rangle$\tablenotemark{a} & Ref. \\
         &  & (s)  & &(deg) &  & ($\,\!''$) & (ms) & ($\%$) & \\
\tableline
 11-01-2012& L27/L\,'+ND & 0.2 $\times$ 80 & 10 & -9.32/-8.19 & 1.39 & 0.75 & 3.6& 61 & \citet{rameau13b} \\
  11-01-2012 & L27/L\,' & 0.2 $\times$ 100 & 156 & -7.59/16.96 & 1.39 & 0.76 & 3.5& 58 & \citet{rameau13b}\\
\tableline
 14-03-2013 & L27/L\,'  & 0.2 $\times$ 100 & 162 &  3.20/28.18  & 1.41 & 1.77 & 1.0& 37 & \citet{rameau13b}\\
 14-03-2013 & L27/L\,'+ND  & 0.2 $\times$80 & 10 &  29.61/30.68  & 1.44 & 1.65 & 0.9 & 32 & \citet{rameau13b}\\
\tableline
 26-06-2013& L27/L\,'+ND & 0.2 $\times$ 80 & 10  &41.0/42.0 & 1.50 & 1.00 & 3.1 & 54 & this work\\
 26-06-2013 & L27/L\,' & 0.2 $\times$ 100 & 96 & 42.5/55.3 & 1.55 & 1.08 &2.8  & 45 & this work\\
 \tableline
  27-06-2013 & L27/L\,' & 0.2 $\times$ 80 & 10 & 28.0/29.1 & 1.44& 1.17 & 1.4& 28  & this work\\
 27-06-2013 & L27/L\,'  & 0.5 $\times$ 100 & 186 & 29.6/58.9& 1.53& 1.02  & 1.6 & 47 & this work \\
\tableline
\end{tabular}
\tablecomments{\scriptsize ``ND'' refers to the NaCo ND\_Long filter (transmission of $\simeq 1.79~\%$),`DIT" to exposure time, and $\pi$ to the parallactic angle at start and end of observations.}
\tablenotetext{a}{\scriptsize The airmass, the seeing $\varpi$, the coherence time $\tau_0 $, and the coherent energy $\mathrm{E}_\mathrm{c}$ are estimated in real time by the adaptive-optics system and averaged here over the observing sequence.}
\end{table*}

To optimize the detection of the faint signal around HD\,95086 and perform high-precision relative astrometry, we observed the star with exactly the same instrumental set-up as in the discovery observations. VLT/NaCo \citep{rousset03, lenzen03} was used in pupil-tracking mode to enable ADI. The observations were carried out at L\,'-band ($\lambda=3.8~\mu m$, $\Delta \lambda=0.62~\mu m$), with the L27 camera ($\simeq27.1$~mas/pixel) with $2\times\mathrm{NDIT}$ (number of frames) short ($0.2~$s) exposures at each of the four-dither position. The faintness of the target as well as the high airmass prevented us from obtaining saturated exposures to achieve higher dynamical range. A short set of unsaturated exposures using a neutral density filter (attenuation of $4.36\pm0.1~$mag) was taken at the beginning of the observing sequence (PSF). The Full-Width-at-Half-Maximum (FWHM) measured on the PSF was of $\sim3.5$~px.

Data were acquired on June-26th and 27th. The observing set up and conditions are detailed in Table \ref{tab:log}\footnote{We also recall the log of the two first datasets in January 2012 and March 2013.}.  On June 26th, although the conditions were stable, the field rotation was $12~$deg, which corresponds to a rotation of only 1.3 FWHM at the expected projected separation of the signal, i.e. $\sim620~$mas. On June 27th, the amplitude of the field rotation was increased ($26.7~$deg) but the conditions were slightly less stable.

The astrometric binary IDS\,1307 \citep{vandessel93} was observed on July 7th in field-tracking mode to calibrate the instrument platescale and orientation. IDS\,1307 was recalibrated on the $\theta_1$ Ori C field (used in January 2012 and March 2013) thanks to contemporaneous observations of both fields obtained in January 2012.

 \subsection{Data reduction and analysis}
 
 In order to avoid systematics or biases from the image processing and ensure a robust detection, five independent pipelines were used to reduce the data. They are described in detail in \citet{boccaletti12} (hereafter LESIA), \citet{currie12} (hereafter A-LOCI), \citet{meshkat13a} (hereafter Leiden), \citet{amara12} ({\sc PynPoint}), and \citet{lagrange10}/\citet{chauvin12} (IPAG-ADI). 
 
Each pipeline processed the data in a similar way for the first steps (flat-fielding, bad/hot pixel removal, sky-subtraction, registration, and frame selection) to create a mastercube of the individual frames together with the list of associated parallactic angles. Several frames affected by bright waffle modes and bright spiders were also rejected. 

The main differences between the pipelines resides in the way the stellar-halo is estimated and subtracted from the mastercube, by using different ADI flavours and different set of parameters. Standard ADI algorithms were applied and we show the results of cADI and sADI \citep{marois06} applied with IPAG-ADI, A-LOCI \citep{currie12} (adapted from \citealt{lafreniere07}), and the most recent PCA-based methods \citep{amara12,soummer12} applied with Leiden, LESIA, and {\sc PynPoint}.
Finally, the residual frames were aligned with the true-North to the vertical and combined by mean averaging. \\
The different ADI techniques produce a high variety of residual images with different speckle intensities and distributions (see Figure \ref{fig:images}). This difference also affects the planet's photometry. Using the different pipelines helped to overcome the possible biases related to each algorithm.
   
In the IPAG-ADI pipeline, the astrometry and photometry were derived as in the discovery paper, following the injection of artificial planets as described in \citet{lagrange10,chauvin12}. In the remaining pipelines, the astrometry was estimated by fitting the planet's signal with a two-dimensional Gaussian or Moffat function. The systematic bias between the two methods was estimated to be less than $0.7~$px using the IPAG pipeline.

The main errors on the position of the companion came from the intrinsic measurement of the source position ($0.4~$px measured by injecting ten artificial planets at the same separation as the companion but different position angles and estimating the effect of the surrounding residuals and procedures.), the star position ($0.1~$px estimated from a series of tests with random shifting of the mastercube and registration as done with the IPAG-ADI pipeline), and finally the astrometric calibration ($0.1~$px). The quadratic sum of the error sources lead to an uncertainty of $0.42~$px. The point-source photometric errors resulted from the uncertainties on the measurements of the signal brightness ($0.7~$mag estimated as for the position), the PSF variability ($0.3~$mag), and the neutral density transmission ($0.1~$mag). Added quadraticaly, we ended up with $0.8~$mag of uncertainty.

The signal-to-noise ratio ($S/N$) was computed following the same approach as in \citet{rameau13b}. The noise-per-pixel was derived from the standard deviation computed in a ring of 1-FWHM width, centered on the star, with a radius equal to the planet-star separation. The planet was also masked within the ring to compute the noise. The flux of the planet was integrated over a 1-FWHM aperture in diameter. The final integrated signal-to-noise ($S/N$) was computed on the same aperture size considering the noise-per-pixel and aperture size in pixels for renormalization. We tested to change the size of both aperture and coronae to check the effect on the $S/N$. All measurement were consistent with the ones from the 1-FWHM adopted value. Although there is probably no optimal way to estimate the S/N in the speckle dominated regime, the same method was applied for each pipeline and each dataset. However, the purpose is not to directly compare the pipelines with their $S/N$ values but rather strengthen the detection.


\section{An unambiguous comoving companion around HD\,95086\label{sec:planet}}

\begin{table*}[t!]
\caption{ADI algorithms and associated parameters on the reduction of the 27-06-2013 data.  \label{tab:adi}}
\centering
\footnotesize
\begin{tabular}{lllllllllll}
\tableline
\tableline
Algorithm & Parameters & S/N & Ref. \\
\tableline
IPAG/sADI & $r=600~$mas, $N_\delta=1~$(FWHM), depth = 6 frames & 7 & \citet{lagrange10} \\
LESIA/PCA & $7$ modes out of 534 & 6.5 & \citet{boccaletti12}\\
Leiden/PCA & $15$ modes out of 185 & 5 & \citet{meshkat13a}\\
A-LOCI & $N_\delta=0.7~$(FWHM), $g=1$, $dr=11$, $N_A=35$, $r_{corr}=0.16$ & 13 & \citet{currie12}\\
{\sc PynPoint}& $40$ coefficients out of $15172$ & 7.5 & \citet{amara12}\\
\tableline
\end{tabular}
\tablecomments{\scriptsize The $S/N$ cannot be used to directly compare each pipeline and reduction algorithm since the distribution and level of the noise is different in each case (For {\sc PynPoint}, the detection was clear and served its main purpose so there was no need to increase the S/N).}
\end{table*}

\subsection{Redetection}

\begin{figure*}[th]
\epsscale{2.}
\plotone{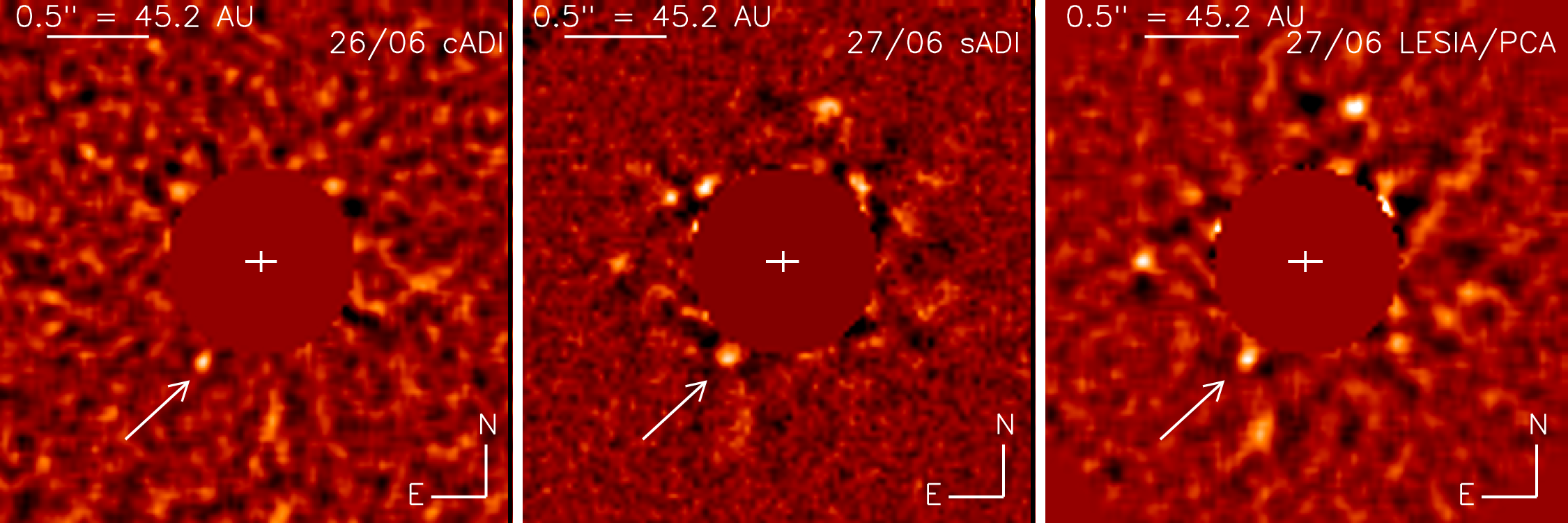}
\plotone{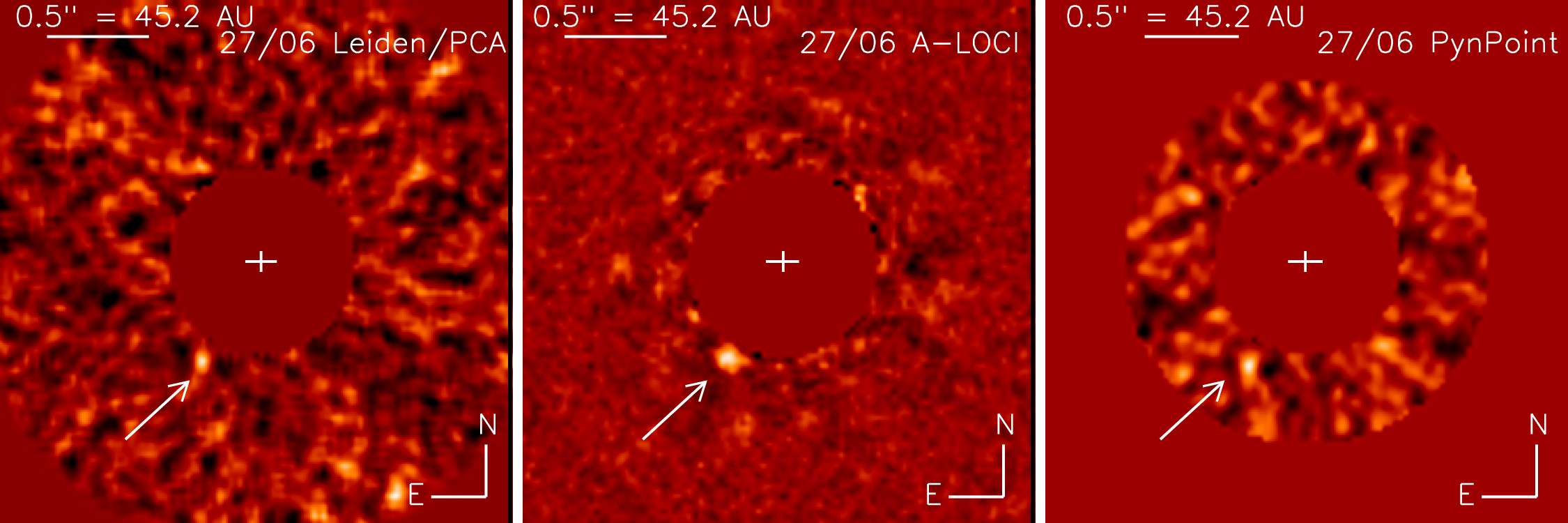}

\caption{Residual maps of VLT/NaCo images at L$\,'$-band, revealing HD\,95086\,b at South-East (arrow). A direct comparison between the pipelines through the $S/N$ is not valid due different noise distributions that are produced. The planet PSF also appears of different sizes due to different cuts and different levels of self-subtraction. \textbf{Top-left:} IPAG-cADI reduction from June 26; $S/N\simeq 4$ due to the small field rotation but good stability. \textbf{Top-central:} IPAG-sADI reduction from June 27;  $S/N\simeq 7$. Speckles at North-East and North-West are strong spike-residuals but at a different separation from the central star than the planet. \textbf{Top-right:} PCA reduction following \citet{boccaletti12} using seven coefficients over 534; $S/N\simeq 6$. \textbf{Bottom-left} Adapted-PCA from \citet{meshkat13a} using 16 coefficients over 185, $S/N\simeq 5$. \textbf{Bottom-central:} A-LOCI from \citet{currie12} where the source is masked over a box of $10~$px in width; $S/N\simeq 13$. \textbf{Bottom-right:} {\sc {\sc Pynpoint}} \citep{amara12} using 40 coefficients over 15172, $S/N\simeq 7.5$. Note {\sc {\sc Pynpoint}} doubles the sampling resolution but it has been rescaled to normal for display purpose.\label{fig:images}}
\end{figure*}

In the 26-06-2013 data, we were able to marginally redetect the signal with all algorithms and pipelines (see cADI example on Figure \ref{fig:images}, top-left, with a $S/N$ between two and four). No other bright point source is seen in the residual maps. Nonetheless, given the low S/N, this dataset alone does not allow a firm confirmation of the planet.

We therefore focus our analysis on the data and outputs from the 27-06-2013 observations. Only this dataset is used in the following analysis. Details of the algorithm parameters are provided in Table \ref{tab:adi}. \\
Figure \ref{fig:images} displays the residual maps with the recovery of HD\,95086\,b. Some speckles have a high intensity but the planet's signal (South-East) is the only one which systematically appears to each pipeline and ADI-flavour. The planet's signal may look different because of first, it is being self-subtracted to different levels and second, different flux levels adopted in each image.

All the five independent pipelines recover the planet's signal at the expected position with $S/N$ higher than five and using the same method for the $S/N$ calculation on the final processed images. The $S/N$ variations between the pipelines are related to the different algorithms used for the PSF subtraction and therefore to the different level and distribution of residuals in the final images. In all cases, the signal is unambiguously detected and confirms the recovery of HD\,95086\,b in our June 2013 data. As an additional check, with the IPAG-ADI pipeline, we injected artificial planets in the raw data at the same separation but different position angles and assuming the brightness measured in the following section. Each fake planet was detected in the residual maps generated with the same $S/N$ as HD\,95086\,b. Moreover, the probability that the signal is a residual speckle is very low since the explored parallactic angles strongly differ from one to another dataset between the different epochs: $[-7.6;17]~$deg in 11-02-2012, $[3.2;28.2]~$deg in 14-04-2013, and $[29.6;58.9]~$deg in 27-06-2013.

Hence, a real, physical object is redetected with a good confidence level from the latest dataset.  
\subsection{Astrometry and photometry}

\begin{table}[th]
\caption{Relative astrometry and photometry of HD\,95086\,b and the background source (bkg star).\label{tab:astro}}             
\centering          
\scriptsize
\begin{tabular}{llll}     
\tableline
\tableline
 Date &  Sep. & PA & $\Delta $L\,' \\
        & (mas) & (deg) & (mag) \\
\tableline
\multicolumn{4}{c}{bkg star}\\
\tableline
 11-01-2012         & $4540\pm15$   & $319.03\pm0.25$    &  $6.2\pm 0.2 $    \\
     14-03-2013      & $4505\pm16$   & $319.42\pm0.26$    &  $6.1\pm 0.2 $    \\
       27-06-2013          & $4480\pm14$   & $319.52\pm0.25$    &  $6.0\pm 0.3 $    \\
   \tableline
   \multicolumn{4}{c}{ HD\,95086\,b}\\
   \tableline
 11-01-2012        & $624\pm 8$   & $151.8 \pm 0.8$    &  $9.79\pm 0.40 $    \\
     14-03-2013           & $626\pm 13$   & $150.7 \pm 1.3$    &  $9.71\pm 0.56 $    \\
    27-06-2013           & $600\pm 11$   & $150.9 \pm 1.2$    &  $9.2\pm 0.8 $    \\
\tableline
\end{tabular}
\end{table}

\begin{figure*}[th]
\epsscale{2.5}
\plottwo{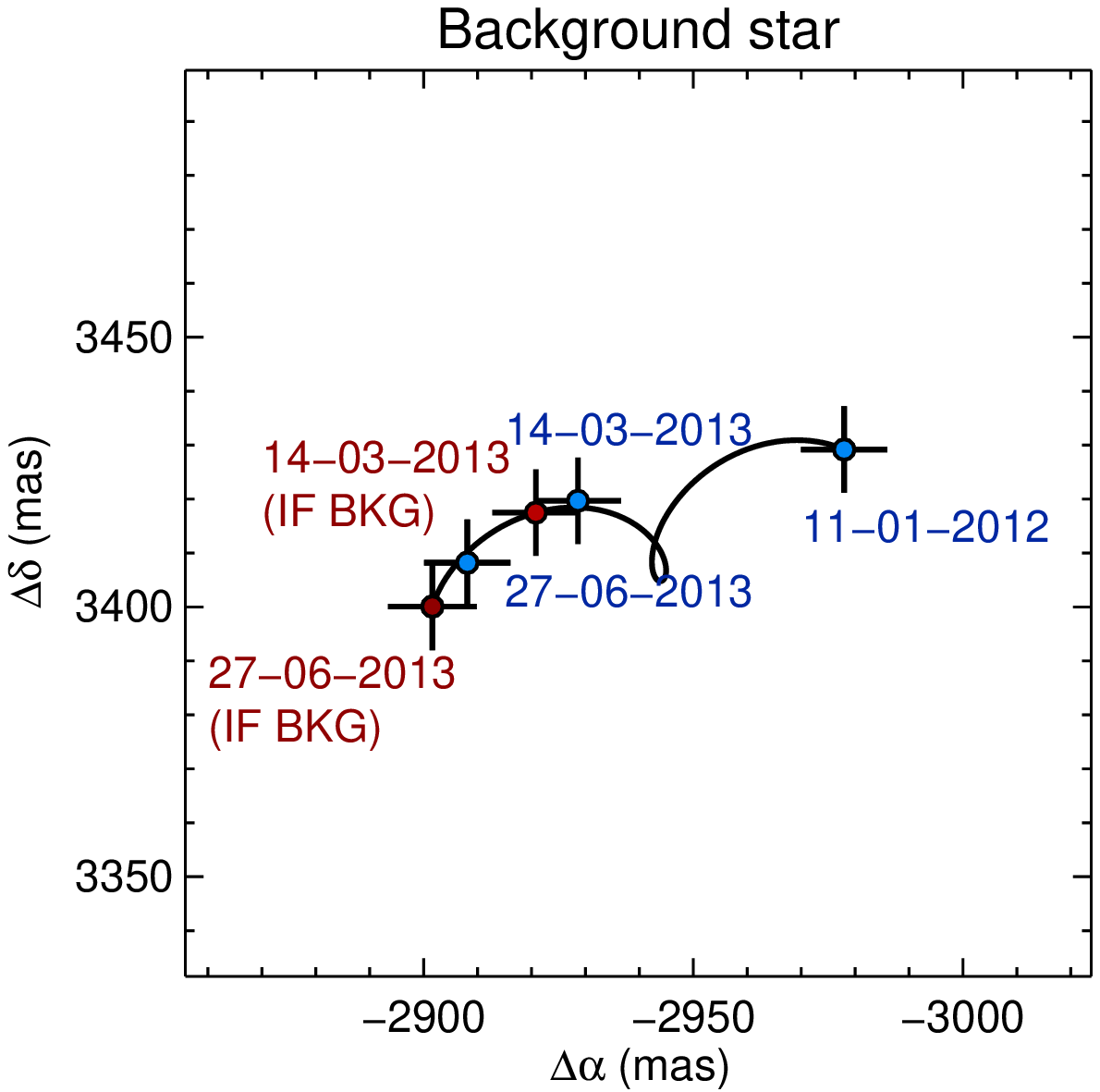}{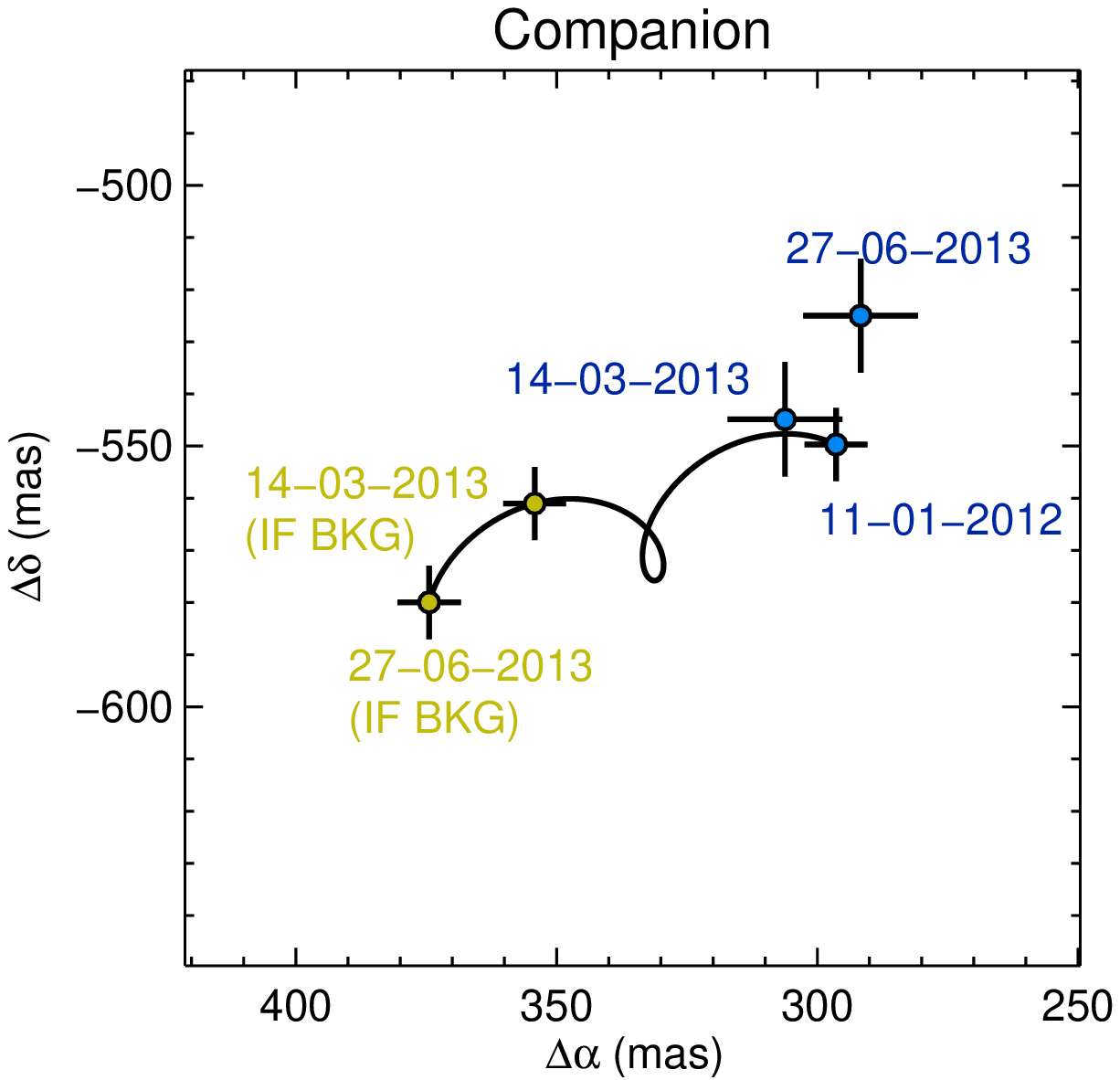}
\caption{Relative separations between the central star and the companion or background source, in right ascension ($\alpha$) and declination ($\delta$). The position measured in 27-06-2013 is over plotted in blue, and the expected position, if the point-source is a fixed background object, is plotted in yellow/red. Previous measurements from \citet{rameau13b} are also reported. \textbf{Left:} Case of the background star. \textbf{Right:} Case of HD\,95086\,b. The 27-06-2013 position lies very close to previous positions and strongly differs from the expected position of a fixed background object. \label{fig:astro}}
\end{figure*}

Since the first two measurements of the probable planet were done within sADI residual maps with the IPAG-ADI pipeline, we used the same reduction algorithm to characterize the present dataset. We estimated a separation of $600\pm 11~$mas and a position angle of $150.9\pm 1.2~$deg. The other pipelines give similar values, all consistent within $20~$mas and $1~$deg.
Figure \ref{fig:astro} shows the relative position of the star with respect to HD\,95086 from each epoch at L\,', and also the track and positions assuming a fixed background object. The position of the companion lies in the same region as the first two epochs and strongly excludes a background object with a $\chi^2$ probability of $10^{-16}$. The point-source is thus comoving with the star.

As done in the discovery paper, we used the $4.5\,''$ background star visible in the field-of-view to ascertain our astrometric measurements. Using Moffat-fitting on co-added (no ADI processing) residual images, we estimated the separation of $4.480\pm0.014\,''$ and a position angle of $319.52\pm0.25~$deg. Our original measurements from January 2012 and March 2013, showing that the point source is a stationary background object, are corroborated by the third point in Figure \ref{fig:astro}., which lies very close to the expected position of June 2013. Therefore, the background star confirms our ability to assess or exclude background behavior with high precision.

The photometric measurement on HD\,95086\,b is $\Delta L\,'=9.2\pm 0.8~$mag, of consistent in all pipelines and with previous estimates.
  

\section{Conclusions and prospects}

New observations of HD\,95086 were carried out with exactly the same instrumental set-up as the discovery paper, namely at L\,'-band with VLT/NaCo in ADI mode. We applied five different, independent pipelines to reduce the data. Each of them led to the confirmation of the point-source to the south-east of the star. Precise astrometric measurements showed the signal is not a background object with a probability of $10^{-16}$. HD\,95086\,b is therefore a companion comoving with its host-star.

From the three set of data, we revised the mass and projected separation of the planet. With $\Delta L\,'=9.79\pm 0.40~$mag in January, 2012, $9.71\pm0.56$ in March, 2013 and $9.2\pm 0.8$ in June, 2013, we estimate its absolute magnitude is $M_L=11.5\pm 1.1$mag. From the BT-Settl models \citep{allard12} which have been shown to be consistent with the red colors of the planet \citep{meshkat13b}, the luminosity corresponds to a mass of $5\pm2$ \Mj~at $17\pm4~$Myr. We remind that ``warm-start'' models \citep{SB12, marleau13} might predict a higher mass (we could only derive a lower limit of $3$~\Mj, see \citealt{rameau13b}). With a predicted effective temperature of $1000\pm200$K, surface gravity of $3.85\pm0.5~$dex, and very red colors, HD\,95086\,b has a cool and dusty atmosphere where the effects of possible non-equilibrium chemistry, reduced surface gravity, and methane bands in the near infrared might be explored in the future. Follow-up observations at different wavelengths, out-of-reach for current facilities, allow characterization of its atmosphere.

With a separation of $623.9\pm 7.4$ mas in January, 2012, $626.1\pm 12.8$ mas in March 2013 and $600\pm 11$ mas in June, 2013, the projected distance to the host-star is $55.7 \pm 2.5$ AU. Recently, \citet{moor13} published resolved \textit{Herschel} images of a debris disk surrounding the star, extending out to $270~$AU, with a possible inclination of about $25~$deg. Based on multi-wavelength observations, they built the spectral energy distribution and adopted, as a best-model, a two-component disk at $6$ and $64~$AU. However, at this stage of modeling, it is not certain whether the disk is best represented by a single or this specific two-component models. If the gap size and positions of the rings are reals, then we can study the properties and influence of HD\,95086\,b on the disk structure. The physical distance of HD\,95086\,b would be about $61.5~$AU, very close to the radius of the outer-cold belt. It might sculpt the inner edge of the belt if its chaotic zone (\citep{wisdom80}) overlaps the belt. Assuming the case of a circular orbit for the planet and planetesimals, \citet{wisdom80} showed that the ring inner edge is located at a separation from the planet $\delta a$ according to the relation $\delta a/a=1.3(M_p/M_s)^{2/7}$, where $a$ is the semi-major axis of the orbit of the planet, $M_p$ its mass, and $M_s$ the host star mass. The relation is satisfied within the uncertainties of all parameters. Therefore, HD\,95086\,b might be responsible for the inner edge of the outer belt. Further orbital monitoring will provide information on the planet's orbital eccentricity and may support this hypothesis if the eccentricity is very small. However, the planet cannot sustain the whole gap alone since its chaotic zone is too small to reach the inner belt at $6~$AU. The presence of additional planets within the gap is required, as for HR\,8799 \citep[e.g.,][]{su09}. In the discovery paper \citep{rameau13b}, we excluded from our sensitivity limits, the presence of any planet more massive than $5~$\Mj beyond about $38~$AU, assuming the inclination of the disk. One might speculate on the number and characteristics of these additional planets from the \citet{wisdom80} relation. Then at least three planets would be needed around $10$, $20$, and $35~$AU, on circular orbits. Further deep observations of the system with next-generation planet imagers might reveal these closer-in planets and further observations of the disk might constrain its physical characteristics. Nevetheless, these are speculations on how to maintain such a wide gap since we do not know neither the true semi-major axis of HD\,95086\,b nor whether the disk has really two separate belts.

Finally, a possible in-situ formation of HD\,95086\,b might be explained with a disk instability scenario \citep{cameron78, boley09, rafikov09} or pebble accretion \citep{lambrechts12}. If formed closer to the star, the presence of more massive companions would be required to excite mean-motion resonances and thus induce {planet-planet scattering \citep{scharf09,veras09}. Again, future deep observations might reveal these additional close-in planets. Another scenario would be planet-disk interaction \citep{papaloizou07, crida09} through outward migration. These scenarios might be tested also through orbital monitoring showing eccentricity and/or structures in more resolved images of the disk.



\acknowledgments
We are grateful to Vanessa Bailey, Kate Su, and Amy Bonsor for fruitful discussions. We thank ESO Director General for discretionary time that allowed to confirm the planet.
 JR, GC, and AML acknowledge financial support from the French National
Research Agency (ANR) through project grant ANR10-BLANC0504-01. 



{\it Facilities:}\facility{VLT: Yepun (NaCo)}.

\end{document}